\documentclass[graybox]{svmult}

\usepackage{mathptmx}       % selects Times Roman as basic font
\usepackage{helvet}         % selects Helvetica as sans-serif font
\usepackage{courier}        % selects Courier as typewriter font
\usepackage{type1cm}        % activate if the above 3 fonts are
\usepackage{makeidx}         % allows index generation
\usepackage{graphicx}        % standard LaTeX graphics tool
\usepackage{multicol}        % used for the two-column index
\usepackage[bottom]{footmisc}% places footnotes at page bottom

\makeindex             % used for the subject index
\begin{document}

\title*{Identification of Pulsation Modes in Main Sequence Stars: Potentials and Limits}
\titlerunning{Identification of Pulsation Modes}
\author{J. Daszy\'nska-Daszkiewicz and A. A. Pamyatnykh}
\institute{J. Daszy\'nska-Daszkiewicz \at Instytut Astronomiczny, Uniwersytet Wroc{\l}awski,
Kopernika 11,  51-622 Wroc{\l}aw, Poland,
\email{daszynska@astro.uni.wroc.pl}
\and A. A. Pamyatnykh \at Copernicus Astronomical Center, Bartycka 18, 00-716 Warsaw, Poland,
\email{alosza@camk.edu.pl}}
\maketitle
% Please use both starred abstract and non-starred abstract.

\abstract*{We review the present-day methods of mode identification
applied to main sequence pulsators focusing on those that make use
of multicolour photometry and radial velocity data. The effects
which may affect diagnostic properties of these observables are
discussed. We also raise the problem of identification of high
$\ell$ modes which can dominate oscillation spectra obtained from
space-based projects.}

\vspace{-8mm}
\abstract{We review the present-day methods of mode identification
applied to main sequence pulsators focusing on those that make use
of multicolour photometry and radial velocity data. The effects
which may affect diagnostic properties of these observables are
discussed. We also raise the problem of identification of high
$\ell$ modes which can dominate oscillation spectra obtained from
space-based projects.}

\vspace{-8mm}
\section{Introduction}
\label{sec:1}

The accuracy of seismic model increases with the number of well
identified pulsation frequencies. Depending on the character of
modes we can probe different parts of a star and derive constraints
on various parameters of model and theory. Since asteroseismology
has entered the space era, the number of detected peaks in the
frequency spectra has grown immensely. However, to explore these
data, the unequivocal identification of mode geometry is required.
In the case of main sequence pulsators this is not an easy task
because their oscillation spectra do not exhibit equidistant or
regular patterns. An alternative is the usage of the photometric and
spectroscopic variations. In this paper, we summarize potentials and
limits in using these observables for mode identification.

In Section\,\ref{sec:2} we recall the photometric diagrams for
$\beta$ Cephei, SPB (Slowly Pulsating B-type) and $\delta$ Scuti
star models. We mention the most important effects which may affect
diagnostic properties of such diagrams, i.e., convection, rotation
and model atmospheres. In Section\,\ref{sec:3} we describe a method
which, beside mode identification, can yield constraints on models
and theory.
%the advantage of adding the radial velocity
A prospect for identification of high $\ell$ modes is discussed in Section\,\ref{sec:4}.
The last Section contains conclusions.

\vspace{-8mm}
\section{Mode identification from multicolour photometry}
\label{sec:2}

\begin{figure}[p]
\includegraphics[clip, width=\textwidth]{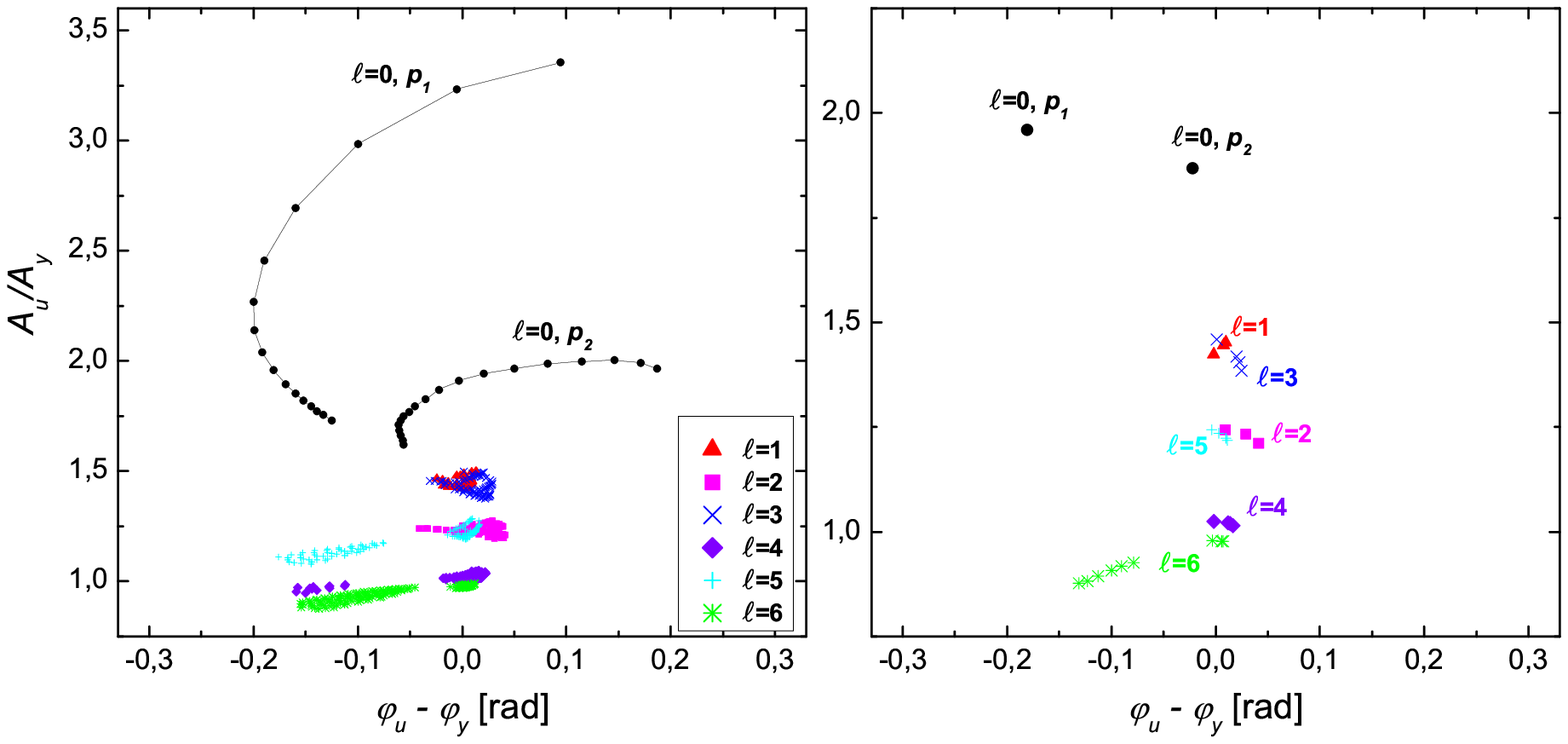}
\includegraphics[clip, width=\textwidth]{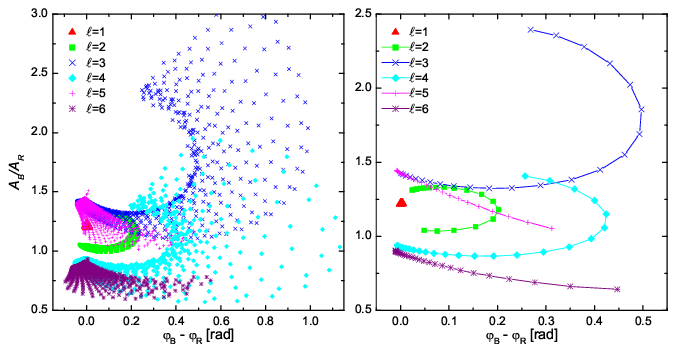}
\includegraphics[clip, width=\textwidth]{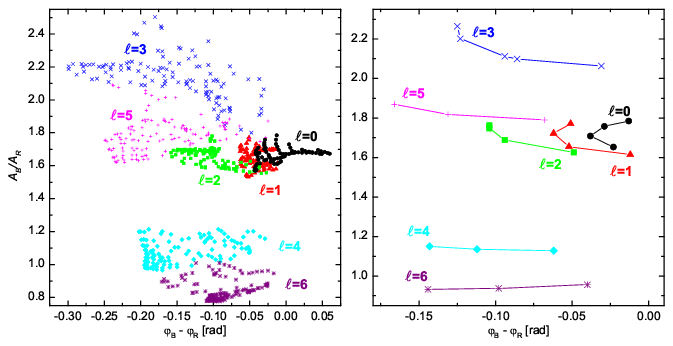}
\caption{The photometric diagnostic diagrams for the main sequence
pulsators: $\beta$ Cep stars (top panels),
SPB stars (middle panels) and $\delta$ Sct stars (bottom panels).
The left panels correspond to all unstable modes
from ZAMS to TAMS for a stellar mass of 12, 5, and 2 $M_\odot$ for $\beta$ Cep,
SPB and $\delta$ Sct models, respectively.
In the right panels unstable modes of particular models are shown. (\cite{JDD2}) }
\label{fig:01}       % Give a unique label
\end{figure}

Since pioneering works \cite{Dziem1}, \cite{Balona}, \cite{Watson},
the photometric observables, i.e., photometric amplitudes and phases
in various passbands, have become the most often used data for mode
identification in main sequence pulsators. If we ignore effects of
rotation, these quantities are independent of the intrinsic mode
amplitude, $\varepsilon$, inclination angle, $i$, and azimuthal
order, $m$. In Fig.\,\ref{fig:01} we show the position of unstable
modes for stellar models with masses of 12, 5 and 2 $M_\odot$
corresponding to $\beta$ Cep, SPB and $\delta$ Sct stars,
respectively. In the case of $\beta$ Cep models we used the
Str\"omgren $uy$ filters and in the case of SPB and $\delta$ Sct
models the Johnson $BR$ filters. The left panels refer to all
unstable modes occurring during the main sequence evolution, whereas
the right panels contain unstable modes for a model with $\log
T_{\rm eff}$=4.400, 4.195 and 3.909 for the $\beta$ Cep, SPB and
$\delta$ Sct model, respectively. We considered modes with the
degree, $\ell$, up to 6.

The position of pulsational modes in the photometric diagrams
changes if effects of rotation are taken into account. This happens
in two cases. The first case is when the frequency difference
between modes is of order of the rotational frequency, and the
degrees, $\ell$, differ by 2. Then, such modes are coupled by
rotation and photometric diagrams become dependent on the
inclination angle and rotational velocity \cite{JDD3}. The second
case is when we deal with slow modes, i.e., modes with frequencies
of order of the rotational frequency. Such modes are typical for the
SPB pulsators. Then, the perturbation approach fails and another
treatment is needed, e.g., traditional approximation. The
photometric diagrams become dependent on $(i,m,V_{\rm rot})$ (e.g.,
\cite{Dziem2}, \cite{JDD7}). In the case of $\delta$ Sct variables,
an additional uncertainty comes from effects of the subphotospheric
convection on pulsation. The use of elaborate model atmospheres
(including also NLTE effects, see \cite{JDD8}) improves the
diagnostic reliability of the photometric diagrams. As an example of
the efficiency of the photometric mode identification combined with
the exact fitting of observed and theoretical frequencies, we note a
study of the $\delta$~Sct variable 44~Tau in which 15 oscillation
frequencies have been detected. For 10 of them the mode degree was
uniquely determined, and the fitting of all 15 frequencies was
achieved giving strong constraints on the stellar model
(\cite{Lenz1}, \cite{Lenz2}).

\vspace{-8mm}
\section{Including radial velocity variations}
\label{sec:3}

\begin{figure}[h]
\includegraphics[clip, width=\textwidth]{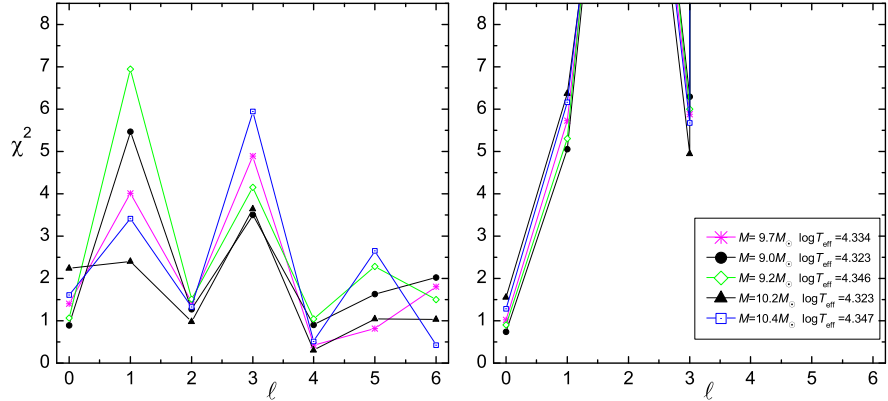}
\caption{ The $\chi^2$ discriminant as a function of $\ell$ for the $\beta$~Cep star $\delta$~Ceti obtained from
the fit of photometric amplitudes and phases without radial velocity
data (left panel) and with radial velocity data (right panel). Different lines correspond to models from
the center and edges of the error box. (\cite{JDD5})}
\label{fig:02}       % Give a unique label
\end{figure}

The uncertainties coming from pulsation theory can be omitted, if
both the mode degree, $\ell$, and the nonadiabatic $f$-parameter are
determined from observations (\cite{JDD4}, \cite{JDD5}). The
$f$-parameter describes the ratio of the radiative flux perturbation
to the radial displacement at the photosphere. Fig.\,\ref{fig:02}
shows results for the $\beta$ Cep star $\delta$~Ceti. In the case of
B-type pulsators, the use of the radial velocity data is essential
to get a unique identification of $\ell$. Moreover, useful
constraints on stellar opacities can be derived.

\begin{figure}[h]
\includegraphics[clip, width=\textwidth]{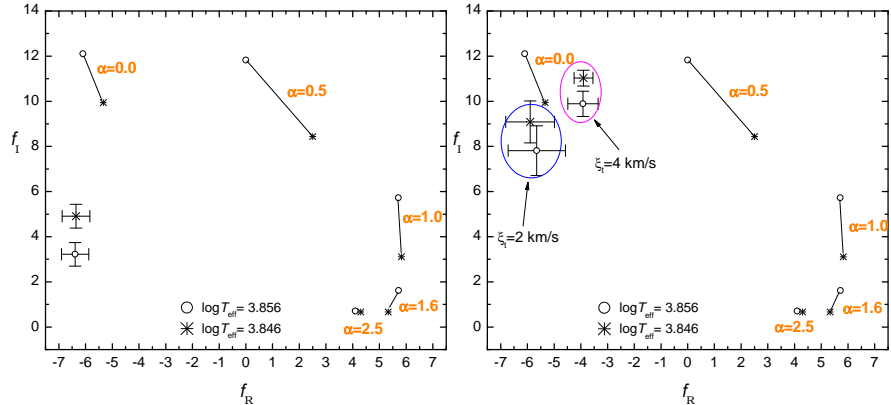}
\caption{Comparison of the empirical values of the $f$-parameter
inferred from Str\"omgren photometry for the $\delta$ Sct star
$\beta$ Cassiopeiae with the theoretical ones calculated for five
values of the MLT parameter, $\alpha$. The empirical values of $f$
were obtained adopting Kurucz models (left panel) and Vienna models
(right panel). In the right panel the effect of the microturbulent
velocity is also shown. (\cite{JDD1})}
\label{fig:03}       % Give a unique label
\end{figure}

Modelling $\delta$ Sct stars is problematic due to the subphotospheric convection
which interacts with pulsation. Mode identification using the method
of simultaneous determination of $\ell$ and $f$ is model independent. On the other hand, a comparison of the theoretical
and empirical values of $f$ provides useful information on the mixing-lenght parameter of convecion
%MLT parameter related to the efficiency of convection
as well as on the model atmospheres. In Fig.\,\ref{fig:03}, we present such a comparison for the $\delta$ Sct star $\beta$ Cassiopeiae
which pulsates in the dipole mode.

\vspace{-8mm}
\section{Prospects for extracting high $\ell$ modes}
\label{sec:4}

The rich oscillation spectra obtained from space observations are
dominated by peaks of very low amplitude. According to the
simulations \cite{JDD6}, there is a high probability that these
peaks are associated to higher $\ell$ modes ($\ell>6$).
\begin{figure}[ht]
\includegraphics[clip, width=\textwidth]{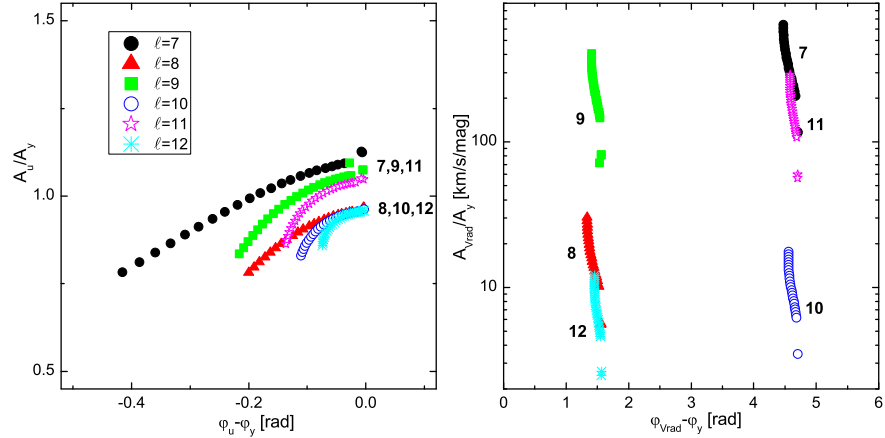}
\caption{Position of modes with $\ell$ from 7 up to 12 in the
diagrams using the Str\"omgren $uy$ passbands (left panel) and the
radial velocity data and the $y$ filter (right panel) for the
$\beta$ Cep model with $M=8.5 M_\odot$ and $\log T_{\rm eff}=4.322$.
Only unstable modes are shown.}
\label{fig:04}       % Give a unique label
\end{figure}

Examples of diagnostic diagrams for a $\beta$ Cep model and
pulsational modes with $\ell$ from 7 up to 12 are shown in
Fig.\,\ref{fig:04}. The left diagram is calculated for the
Str\"omgren $uy$ passbands. The amplitude ratio for any two
photometric passbands of the high $\ell$ modes is around 1, because
the geometrical effect, $(1-\ell)(\ell+2)$, dominates in the light
variations. As we can see there is some mode grouping. This is
related to the parity of degree, $\ell$. The odd $\ell$ modes have
higher values of the amplitude ratio $A_u/A_y$ than the even $\ell$
modes. The right diagram shows the same modes but using radial
velocity variations and the $y$ filter. Here, the mode grouping is
twofold. First, the phase differences take the values according to
the sign of the disc averaging factor, $b_\ell$. Modes with the
phase difference value $\varphi_{Vrad}-\varphi_y$ around 1.5 rad
correspond to negative values of $b_\ell$, whereas those with
$\varphi_{Vrad}-\varphi_y$ around 4.6 correspond to positive values
of $b_\ell$. Second, the values of the amplitude ratio, $A_{\rm
Vrad}/A_y$, depend on the parity of $\ell$. The even $\ell$ modes
have $A_{\rm Vrad}/A_y < 30$ whereas the odd $\ell$ modes have
$A_{\rm Vrad}/A_y >50$. In both panels of Fig.\,\ref{fig:04}, there
is a separation of p and g-modes. The g-modes have negative values
of $\varphi_u-\varphi_y$ and larger values of $A_{\rm Vrad}/A_y$.

For much higher $\ell$ mode,  $\ell>20$, the amplitude ratio for any
two passbands will be around 1, and the phase difference around 0
for all modes. The amplitude ratios for the photometric passband and
radial velocity will be spread from 0 to  about 100 and the phase
differences will have values of 1.5 or 4.6 rad.

\vspace{-8mm}
\section{Conclusions}
\label{sec:5}
\vspace{-3mm}

In this short review, we mentioned the basic properties and
uncertainties of mode identification from multicolour photometry and
radial velocity data. We focused on main sequence pulsators because
of their irregular oscillation spectra.

Then, we checked a prospect for identification of the high degree
modes. Such modes are highly expected in the rich frequency spectra
obtained from, e.g., the CoRoT, Kepler and BRITE data.
Identification of the high $\ell$ modes only from multicolour
photometry is rather hopeless. However, it turned out that it is
possible to get some constraints on intermediate $\ell$ modes from
the diagrams using amplitudes and phases of the photometric and
radial velocity changes. Unfortunately, there is no prospect for
extracting modes with $\ell>30$ from these data. Here only the
analysis of the line-profile variations may help.

%\begin{acknowledgement}
%If you want to include acknowledgments of assistance and the like at the end of an individual chapter please use the \verb|acknowledgement| environment -- it will automatically render Springer's preferred layout.
%\end{acknowledgement}

%\vspace{-1mm}
%
\begin{acknowledgement}
The authors acknowledge partial financial support from the Polish MNiSW grant No. N N203 379 636.
\end{acknowledgement}

\vspace{-12mm}

\end{document}